\documentclass[final,3p,times]{elsarticle}
\usepackage{amsfonts}
\usepackage{mathrsfs}
\usepackage{amsmath}
\usepackage{xcolor}
\usepackage{amssymb}
\usepackage{bm}

\usepackage[linktocpage=true,bookmarks=true,bookmarksnumbered=true,breaklinks=true,pdfpagemode=Fullscreen,pdfstartview=FitBH]{hyperref}

\renewcommand{\rm}{\mathrm}
\def\bge{\begin{equation}}
\def\ede{\end{equation}}
\def\bga{\begin{aligned}}
\def\eda{\end{aligned}}
\newcommand{\beq}{\begin{equation}}
\newcommand{\eeq}{\end{equation}}
\newcommand{\bq}{\begin{equation}}
\newcommand{\eq}{\end{equation}}
\newcommand{\ba}{\begin{array}}
\newcommand{\ea}{\end{array}}
\newcommand{\beqa}{\begin{eqnarray}}
\newcommand{\eeqa}{\end{eqnarray}}
\newcommand{\beqs}{\begin{subequations}}
\newcommand{\eeqs}{\end{subequations}}

\def\dis{\displaystyle}

\def\({\left(}
\def\){\right)}
\def\[{\left[}
\def\]{\right]}

\def\End{\end{document}}

\def\geqq{\geqslant}

\def\dif{\partial}
\def\al{\alpha}
\def\be{\beta}

\def\to{\rightarrow}

\def\ii{\mathrm{i}}

\def\ii{\textrm{i}}

\def\geqq{\geqslant}

\def\dif{\partial}
\def\al{\alpha}
\def\be{\beta}

\def\End{\end{document}}

\begin{document}

\begin{frontmatter}

\setcounter{footnote}{1}
\renewcommand{\thefootnote}{\fnsymbol{footnote}}

\title{{\bf Origin of Constrained Maximal CP Violation in Flavor Symmetry}}

\author{
{\sc Hong-Jian He}\,\footnote{hjhe@tsinghua.edu.cn}\,$^{a,b}$,~~
{\sc Werner Rodejohann}\,\footnote{werner.rodejohann@mpi-hd.mpg.de}\,$^{c}$,~~
{\sc Xun-Jie Xu}\,\footnote{xunjie.xu@gmail.com}\,$^{a,c}$
\vspace*{3mm}
}

\address{
$^a$\,Institute of Modern Physics and Center for High Energy Physics,
Tsinghua University, Beijing 100084, China
\\[1mm]
$^b$\,Institute for Advanced Study, Princeton, NJ 08540, USA
\\[1mm]
$^c$\,Max-Planck-Institut f\"{u}r Kernphysik, Postfach 103980, D-69029 Heidelberg, Germany
}

\begin{abstract}
Current data from neutrino oscillation experiments are in good agreement with
$\,\delta=-\frac{\pi}{2}\,$ and $\,\theta_{23}^{} = \frac{\pi}{4}$\, under the
standard parametrization of the mixing matrix.
We define the notion of \,``constrained maximal CP violation''
(CMCPV) for predicting these features and study their origin in flavor symmetry.
We derive the parametrization-independent solution of CMCPV and
give a set of equivalent definitions for it.
We further present a theorem on how the CMCPV can be realized.
This theorem takes the advantage of residual symmetries in neutrino and
charged lepton mass matrices, and states that, up to a few minor exceptions,
$\,(|\delta |,\,\theta_{23}^{}) =(\frac{\pi}{2},\,\frac{\pi}{4})\,$
is generated when those symmetries are real. The often considered $\mu$-$\tau$ reflection
symmetry, as well as specific discrete subgroups of O(3), are special cases of our theorem.
\\[1.5mm]
\\[1.5mm]
PACS numbers: 11.30.Hv, 12.15.Ff, 14.60.Pq  \hfill
Phys.\ Lett.\ B (2015), in Press $[${arXiv:1507.03541}$]$
\end{abstract}


\end{frontmatter}



\setcounter{footnote}{0}
\renewcommand{\thefootnote}{\arabic{footnote}}

\section{Introduction}
\label{sec:1}

While a coherent picture in leptonic mixing has emerged,
important measurements are still lacking. In particular,
the Dirac CP angle $\,\delta\,$ and the exact value of the atmospheric neutrino
mixing angle $\,\theta_{23}^{}\,$ are of great interest.
Whether $\,\theta_{23}^{}\,$ is maximal or departs sizably from $\,\frac{\pi}{4}\,$
has important ramifications for flavor symmetry models \cite{Altarelli:2010gt}. The CP phase also
has model building impact, and the question of whether the lepton sector violates CP has
conceptual significance in connection to the matter-antimatter asymmetry via
leptogenesis \cite{Davidson:2008bu}.  While maximal atmospheric mixing is compatible with data
since the observation of atmospheric neutrino oscillations,
recently first hints towards a Dirac CP angle $\,\delta=-\frac{\pi}{2}\,$
have arisen from the appearance and disappearance measurements of T2K \cite{T2K:2013hdq}
when combined with reactor antineutrino data.  Indeed, global fits \cite{NeuFitLisi,Forero:2014bxa,Elevant:2015ska} confirm a mild preference
for this particular value of CP phase.

With these in mind, it is tempting to study the origin of such values
of $\,\delta\,$ and $\,\theta_{23}^{}\,$ within theories of flavor symmetry.
In particular, the so-called
$\mu$-$\tau$ reflection symmetry\,\cite{Harrison:2002et,Ma:2002ce,Grimus:2003yn,MR}
was often considered in the literature in this respect. It transforms the neutrino fields as
$\,(\nu_e^{},\,\nu_\mu^{},\,\nu_\tau^{})\to
   (\nu_e^*,\,\nu_\tau^*,\,\nu_\mu^*)$\,,\, leading to
$\,|\delta|=\frac{\pi}{2}\,$ and $\,\theta_{23}^{}=\frac{\pi}{4}\,$
in the standard parametrization of the PMNS mixing matrix\,\cite{PMNS,PDG}.
In our study, we demonstrate that these two features arise as the outcome
of ``Constrained Maximal CP Violation'' (CMCPV), which we will establish
{\it in a parametrization-independent way by maximizing the Jarlskog invariant
under a minimal constraint.}

The framework we will discuss is that a flavor symmetry group $\,G\,$ is broken such
that the neutrino and charged lepton mass matrices are invariant
under certain subgroups of $\,G$\,.\,
We will propose and prove a general theorem revealing that
if the residual flavor symmetries are real\footnote{Here and henceforth
``a symmetry is real'' always means that the transformation matrix
representing the symmetry is real.},\, then the CMCPV is generated.
There are a few minor exceptions to this theorem which we will clarify in Sec.\,\ref{sec:3}.
The $\mu$-$\tau$ reflection symmetry is actually a special case of this
theorem, which can be shown explicitly after a simple basis transformation.
We further deduce some corollaries from the theorem which
are practically useful in understanding and building models for the CMCPV.
For instance, specific subgroups of O(3) can generate CMCPV,
so do the models with certain groups under which all neutrino fields transform
as triplets. As an illustration, we will present a simple model to explicitly
realize the CMCPV.

This paper is organized as follows. In Sec.\,\ref{sec:parametrization},
we will establish our definition of CMCPV in a parametrization-independent way,
and give a set of equivalent descriptions. Various physical implications
(such as leptonic unitarity triangles) from CMCPV are further discussed.
In Sec.\,\ref{sec:A-Theorem-for}, we present our theorem for the origin of
CMCPV and derive its corollaries which are important for practical model buildings.
We will study applications in Sec.\,\ref{sec:Application},
and finally we conclude in Sec.\,\ref{sec:con}. Some elaborated mathematical
proofs are presented in Appendices A and B.

\vspace*{2mm}
\section{Parametrization-Independent Formulation of Constrained Maximal CP Violation}
\label{sec:parametrization}
\label{sec:2}

What we mean by ``constrained maximal CP violation'' (CMCPV) is not merely
$\,|\delta| = \frac{\pi}{2}\,$
in the standard parametrization of the PMNS matrix\,\cite{PMNS,PDG},
but both the $\,|\delta| = \frac{\pi}{2}\,$
and $\,\theta_{23}^{} = \frac{\pi}{4}\,$.\,
In general, a parametrization-independent definition of the maximal CP violation should
be given in terms of Jarlskog invariant $J$ \cite{J}, rather than the CP
angle $\,\delta\,$,\, because $\,\delta\,$ is not rephasing invariant. Furthermore, we will
clarify shortly that naively maximizing $\,J\,$ without constraint is already
excluded by experimental data.
Hence, introducing the {\it new concept of CMCPV} is essential for studying the
viable maximal CP violation.
For this purpose, we first formulate the CMCPV in a parametrization-independent form.
\\[2mm]
\noindent
{\bf Definition\,1}\,(CMCPV):
\\
\emph{Constrained Maximal CP Violation (CMCPV) is defined as the maximum of the
absolute value of Jarlskog invariant under the minimal constraint
that the absolute values of the elements in the first row of the PMNS matrix
are fixed.}
\\[-1.5mm]

The Definition\,1 is parametrization-independent because it does not invoke any
explicit form of the PMNS matrix.
Note that this is a constrained maximization problem. The Jarlskog
invariant $J$ \cite{J} can be regarded as a function of the PMNS matrix $U$.
We are looking for the maximal values of the function $\,J(U)$\,,\, where
$\,U\,$ is not an arbitrary unitary matrix but a constrained one.  We impose
this constraint on $\,U\,$
by requiring the absolute values of its elements in the first row,
$\,(|U_{e1}|,\,|U_{e2}|,\,|U_{e3}|)$,\, be fixed to certain given values.
(Actually, fixing any two of them in the first row is enough,
due to the unitarity condition
$\,|U_{e1}|^2\!+|U_{e2}|^2\!+|U_{e3}|^2=1$.)\,
The reason that we fix absolute values of the elements of $U$ in its first row,
rather than any other rows or columns, will become clear shortly
(cf.\ footnote-2).

This constraint is necessary because without it $\,J\,$ would reach its
maximal value as allowed by unitarity, $|J| = \frac{1}{6\sqrt{3}}$,
which is equivalent to all unitarity
triangles being equilateral. The corresponding $\,U\,$ in
this case is just the Wolfenstein mixing matrix \cite{Wolfenstein:1978uw},
\begin{equation}
U_{W}^{} =\, \frac{1}{\sqrt{3}\,}
\left(
\begin{array}{ccc}
1 & 1 & 1\\
1 & \,\omega^{2} & \omega\\
1 & \omega & \,\omega^{2}
\end{array}
\right) ,
\label{eq:0621}
\end{equation}
with $\,\omega = e^{\ii 2\pi /3}$,\,  which has been excluded by oscillation data.
In the standard parametrization\,\cite{PDG}, the Jarlskog invariant is given by
\beqa
J \,=\,
\frac{1}{8}\sin\delta\cos\theta_{13}\sin2\theta_{13}\sin2\theta_{23}\sin2\theta_{12}\,.
\eeqa
Indeed, if we compute its maximum by $\,\partial_{\theta}^{}J=0$\,
with $\,\theta=\{\theta_{12}^{},\,\theta_{13}^{},\,\theta_{23}^{},\,\delta\}$,\,
we obtain the Wolfenstein mixing,
$\,(\theta_{13}^{},\,\theta_{12}^{},\,\theta_{23}^{},\,|\delta|)
 = \(\arctan\!\frac{1}{\!\sqrt{2\,}\,},\,\frac{\pi}{4},\,\frac{\pi}{4},\,\frac{\pi}{2}\)\,$.\,
This includes the desired values of $\,(\theta_{23}^{},\,\delta)$,\,
but gives unrealistic  $\,(\theta_{12}^{},\,\theta_{13}^{})$.\,
Hence, the Wolfenstein mixing is already excluded by experimental data.
To derive acceptable maximization of $\,J$\,,\,
we observe that if we fix $\,\theta_{12}^{}$\, and \,$\theta_{13}^{}$\,
(to their best-fit values for instance) and
then maximize $\,J\,$,\, we still obtain
$(|\delta|,\,\theta_{23}^{}) = \(\frac{\pi}{2},\,\frac{\pi}{4}\)$\,.\,
This is in fact consistent with the above
{\it parametrization-independent Definition\,1 of CMCPV\,,} because
fixing $\,(\theta_{12}^{},\,\theta_{13}^{})$\, corresponds to fixing
the absolute values of the elements in the first row of $\,U$\,
under the standard parametrization.
Note that this is the allowed  minimal constraint we could impose on the Jarlskog invariant:
fixing any other row or column of the PMNS matrix
and then maximizing $\,J\,$ will not lead to
experimentally acceptable results.\footnote{To be explicit, we have directly verified
that fixing the second row or the third row of the PMNS matrix
will result in $\,|U_{ej}^{}|=|U_{\tau j}^{}|$\,,\, or
$\,|U_{ej}^{}|=|U_{\mu j}^{}|$\,,\, ($j=1,2,3$), respectively.
Fixing the first, second or third columns will lead to
$\,|U_{\ell 2}^{}|=|U_{\ell 3}^{}|$\,,\,
$|U_{\ell 1}^{}|=|U_{\ell 3}^{}|$\,,\, or
\,$|U_{\ell 1}^{}|=|U_{\ell 2}^{}|$\,,\,
($\ell =e,\mu,\tau$),\, respectively.
All these cases are already excluded by the current neutrino oscillation data.}\,
Hence, the above Definition\,1 gives a minimal parametrization-independent definition of
viable maximal CP violation.

It is worth noting that
the Jarlskog invariant can be further cast into a manifestly
parametrization-independent form
\cite{J2},
\beqs
\label{eq:J2-all}
\beqa
J^{2}
&=&
\left|U_{\al i}^{}\right|^{2}\!\left|U_{\al j}^{}\right|^{2}\!
\left|U_{\be i}^{}\right|^{2}\!\left|U_{\be j}^{}\right|^{2}\!
-\frac{1}{4}\left(1\!
-\left|U_{\al i}^{}\right|^{2}\!\! - \left|U_{\al j}^{}\right|^{2}\!\!
-\left|U_{\be i}^{}\right|^{2}\!\! - \left|U_{\be j}^{}\right|^{2}\!\!
+\left|U_{\al i}^{}\right|^{2}\!\left|U_{\be j}^{}\right|^{2}\!\!
+\left|U_{\al j}^{}\right|^{2}\!\left|U_{\be i}^{}\right|^{2}
\right)^{2}
\label{eq:J2}
\\
&=&
\left|U_{e1}^{}\right|^{2}\!\left|U_{e3}^{}\right|^{2}\!
\left|U_{\mu 1}^{}\right|^{2}\!\left|U_{\mu 3}^{}\right|^{2}\!
-\frac{1}{4}\left(1\!
-\left|U_{e1}^{}\right|^{2}\!\! - \left|U_{e3}^{}\right|^{2}\!\!
-\left|U_{\mu 1}^{}\right|^{2}\!\! - \left|U_{\mu 3}^{}\right|^{2}\!\!
+\left|U_{e1}^{}\right|^{2}\!\left|U_{\mu 3}^{}\right|^{2}\!\!
+\left|U_{e3}^{}\right|^{2}\!\left|U_{\mu 1}^{}\right|^{2}
\right)^{2}\!.
\hspace*{10mm}
\label{eq:J2-emu13}
\eeqa
\eeqs
The above Eq.\,\eqref{eq:J2} is a general expression with $\,(\al,\be )=e,\mu,\tau$\,
and $\,(i,j)=1,2,3$,\, where $\,\al\neq\be\,$ and $\,i\neq j\,$.\,
The Eq.\,\eqref{eq:J2-emu13} corresponds to the case of
$\,(\al,\,\be )=(e,\,\mu)$\, and $\,(i,\,j)=(2,\,3)$.\,
It is clear that the Jarlskog invariant can be fully determined by
any 4 independent matrix elements
$\,\{|U_{\al i}|,\,|U_{\al j}|,\,|U_{\be i}|,\,|U_{\be j}|\}$\,
with $\,\al\neq\be\,$ and $\,i\neq j\,$,\,
as shown in Eq.\,\eqref{eq:J2}.
According to our above Definition\,1 of CMCPV and using Eq.\,\eqref{eq:J2-emu13},
we can maximize Jarlskog invariant
by imposing the extremal conditions with the two elements
$(|U_{e1}|,\,|U_{e3}|)$ of first row fixed,
$\,\partial J^2/\partial |U_{\mu 1}|^2 =0\,$ and
$\,\partial J^2/\partial |U_{\mu 3}|^2 =0\,$.\,
From these two equations, we can derive the solution of CMCPV,
\beqa
\left|U_{\mu j}^{}\right|^2  \,=\,
\left|U_{\tau j}^{}\right|^2 \,=\,
\frac{1}{2}\(1-|U_{ej}^{}|^2\)
, \hspace*{8mm} (j=1,2,3)\,,
\label{eq:|U2j|=|U3j|}
\eeqa
which is the maximum of $\,J\,$ under the minimal constraint of Definition\,1,
and is manifestly parametrization-independent.
The detail of this derivation is presented in \ref{App:A}.

When adopting the standard parametrization\,\cite{PDG} of PMNS mixing matrix $U$,\,
we can use Eq.\,\eqref{eq:|U2j|=|U3j|} to immediately deduce
the explicit realization of CMCPV,
$\,(|\delta|,\,\theta_{23}^{}) = \(\frac{\pi}{2},\,\frac{\pi}{4}\)$\,,\,
which is proven in \ref{App:A}.
Furthermore, we find that once we realize
$\,(|\delta|,\,\theta_{23}^{}) = \(\frac{\pi}{2},\,\frac{\pi}{4}\)$\,
under the standard parametrization,
the PMNS matrix exhibits an interesting feature, which we explain as follows.
The standard parametrization of the PMNS matrix is expressed as\,\cite{PDG},
{\small
\beq
\hspace{-1mm}
U \,=\, \left(\!\!\begin{array}{ccl}
c_{12}^{}c_{13}^{} & s_{12}^{}c_{13}^{} & s_{13}^{}e^{-\ii\delta}
\\[0.8mm]
-s_{12}^{}c_{23}^{} \!- c_{12}^{}s_{23}^{}s_{13}^{}e^{\ii\delta}
& c_{12}^{}c_{23}^{}\!- s_{12}^{}s_{23}^{}s_{13}^{}e^{\ii\delta}
& s_{23}^{}c_{13}^{}
\\[0.8mm]
s_{12}^{}s_{23}^{}\! - c_{12}^{}c_{23}^{}s_{13}^{}e^{\ii\delta}
& -c_{12}^{}s_{23}^{}\! -c_{23}^{}s_{12}^{}s_{13}^{}e^{\ii\delta}
& c_{23}^{}c_{13}^{}
\end{array}\!\!\right)  ,~~~~~
\label{eq:0623}
\eeq
}
\hspace*{-2.5mm}
where we have used the common notations
$(s_{ij}^{},\,c_{ij}^{})=(\sin\theta_{ij}^{},\,\cos\theta_{ij}^{})$.\,
Under a rephasing
\beqa
U' \,\equiv\,
\textrm{diag}(1,\,e^{-\ii\delta},\,e^{-\ii\delta})\,U\,
\textrm{diag}(1,\,1,\,e^{\ii\delta})\,,~~~~~
\label{eq:0623-1}
\eeqa
we obtain
{\small
\beqa
\hspace{-3mm} U' \,=\, \left(\!\!\begin{array}{ccc}
c_{12}^{}c_{13}^{} & s_{12}^{}c_{13}^{} & s_{13}^{}
\\[0.8mm]
-c_{12}^{}s_{23}^{}s_{13}^{}\!- s_{12}^{}c_{23}^{}e^{-\ii\delta}
& -s_{12}^{}s_{23}^{}s_{13}^{}\!+ c_{12}^{}c_{23}^{}e^{-\ii\delta}
& s_{23}^{}c_{13}^{}
\\[0.8mm]
-c_{12}^{}c_{23}^{}s_{13}^{} \!+ s_{12}^{}s_{23}^{}e^{-\ii\delta}
& -s_{12}^{}c_{23}^{}s_{13}^{} \!- c_{12}^{}s_{23}^{}e^{-\ii\delta}
& c_{23}^{}c_{13}^{}
\end{array}\!\!\right) .
\nonumber
\\[-1mm]
\label{eq:0623-2}
\eeqa
}
For $\,(\delta,\,\theta_{23}^{}) = \(\pm\frac{\pi}{2},\,\frac{\pi}{4}\)$,\,
we find that $\,U'\,$ becomes
{\small
\begin{equation}
U'_{\!m} \,=\, \frac{1}{\sqrt{2\,}\,}\left(\!\!
\begin{array}{ccc}
\sqrt{2}c_{12}^{}c_{13}^{} & \sqrt{2}s_{12}^{}c_{13}^{} & \sqrt{2}s_{13}^{}
\\[1mm]
-c_{12}^{}s_{13}^{}\pm \ii\hspace*{0.3mm} s_{12}^{}
& -s_{12}^{}s_{13}^{} \mp \ii\hspace*{0.3mm} c_{12}^{} & c_{13}^{}
\\[1mm]
-c_{12}^{}s_{13}^{}\mp \ii\hspace*{0.3mm} s_{12}^{}
& -s_{12}^{}s_{13}^{} \pm \ii\hspace*{0.3mm} c_{12}^{} & c_{13}^{}
\end{array}\!\!\right) .~~~
\label{eq:0623-3}
\end{equation}
}
Eq.\,\eqref{eq:0623-3} explicitly reveals that the first row is real,
while the second row and the third row are complex conjugates of each other.
We call this feature the ``row conjugation equality'' (RCE).
It is easy to see that the reverse is also
true: holding RCE will result in
$\,(|\delta|,\,\theta_{23}^{}) = \(\frac{\pi}{2},\,\frac{\pi}{4}\)$\,.\,
Even though we have demonstrated RCE by using the standard parametrization
\eqref{eq:0623} \cite{PDG},
we stress that the RCE form should be independent of parametrizations (up to trivial rephasing).
We can use any other parametrizations\,\cite{Kobayashi:1973fv}\cite{Fritzsch:1997fw}
and maximize Jarlskog invariant under the same constraint as in Definition\,1.
Then, we find that the mixing matrix always exhibits the RCE form,
up to a trivial rephasing. In fact, we see that any specific RCE form
does obey $\,|U_{\mu j}^{}| = |U_{\tau j}^{}|\,$
as in our parametrization-independent general solution \eqref{eq:|U2j|=|U3j|} of the CMCPV.

\vspace*{1mm}
For later usage, let us introduce the following lemma on the RCE.
\\[2mm]
\noindent
{\bf Lemma}~$[\,$O(3) invariance of RCE\,$]$:
\\
\emph{If a unitary matrix $\,V$\, has the form of RCE, then after a right-handed
real transformation $\,V\to V'=VR$\,,\,  the matrix $\,V'$ should still have the form of RCE,
where $\,R\in \text{O(3)}\,$ is an arbitrary orthogonal matrix.}\\[2mm]
The proof of this statement is delegated to \ref{sec:a3}.

\vspace*{1.5mm}

Another interesting feature of RCE concerns leptonic unitarity triangles (LUTs)
of the PMNS matrix $\,U$,\, in connection to its column orthogonality,
\beqa
U_{ei}^{*}U_{ej}^{} + U_{\mu i}^{*}U_{\mu j}^{}
+ U_{\tau i}^{*}U_{\tau j}^{} =\, 0\,,
\label{eq:0707-7}
\eeqa
where the column indices $\,i\neq j\,$.\,
We call these unitarity triangles the \emph{column triangles.}
It is evident that if \,$U$\, has the form of RCE, then all column triangles
should be \emph{isosceles triangles} because under the RCE the two sides
$\,U_{\mu i}^{*}U_{\mu j}^{}\,$ and
$\,U_{\tau i}^{*}U_{\tau j}^{}$\, have equal length,
$\,|U_{\mu i}^{*}U_{\mu j}^{}|=|U_{\tau i}^{*}U_{\tau j}^{}|$\,.

Unitarity triangles are intrinsically connected to CP violation because all these
triangles have the same area, which equals half of the absolute value of
Jarlskog invariant $\,J$.\,
The LUTs are less studied than the unitarity triangles in quark sector
since measuring the LUTs and thus the leptonic CP violation is much harder.
Nevertheless, the LUTs can be directly measured in principle
via oscillation experiments \cite{He:2013rba}.
Furthermore, the LUTs can provide a geometrical formulation of the CMCPV.
Since we define CMCPV as $\,J\,$ reaching its maximum under certain constraints,
it also means that the area of the LUT reaches its maximum under those constraints.
How do these constraints appear in our current geometrical picture?
The constraint in our Definition\,1 is that the first row of $\,U\,$ is fixed,
namely, \,$|U_{e1}^{*}U_{e2}^{}|$,\, $|U_{e2}^{*}U_{e3}^{}|$\, and \,$|U_{e3}^{*}U_{e1}^{}|$\,
are fixed, which means that the $e$-sides of the column triangles are fixed.
Hence, the Definition\,1 is equivalent to saying that each column triangle
reaches its maximal area with its $e$-side fixed. This provides a
{\it geometrical formulation} of the CMCPV. 

Note that for a triangle with its $e$-side fixed and its perimeter
(the sum of the lengths of its three sides) bounded from above,
its area reaches the maximum if and only if it is an isosceles triangle.
This is clear from geometrical intuition.
In Ref.\,\cite{Xu:2014via}, we proved
that a unitarity triangle must always have its perimeter equal or
less than $1$.\,  This is a necessary and sufficient condition for
a triangle to be unitarity triangle, and requires the perimeter of
each unitarity triangle to be bounded from above, which ensures the area
of each unitarity triangle to have a maximum.
With these, we give a geometrical formulation of the CMCPV:
it corresponds to the maximal area of the LUT by fixing its $e$-side,
and such LUT is an isosceles triangle.

Finally, we summarize the analysis of this section into the following theorem.
\\[2mm]
\noindent{\bf Theorem\,1}~[Equivalent definitions of CMCPV].
\\[0.5mm]
{\it For the PMNS mixing matrix \,$U$,\, the following statements are equivalent:}
\begin{itemize}

\vspace*{-2mm}
\item[(a).]{\it it has the CMCPV (cf.\ Definition\,1);}

\vspace*{-2mm}
\item[(b).]{\it for any parametrization of \,$U$,\,
the general condition \eqref{eq:|U2j|=|U3j|} holds;}

\vspace*{-2mm}
\item[(c).]{\it in the standard parametrization,
$(|\delta|,\,\theta_{23}^{}) = \(\frac{\pi}{2},\,\frac{\pi}{4}\)$\, holds;}

\vspace*{-2mm}
\item[(d).]{\it it has the form of  RCE (up to rephasing);}

\vspace*{-2mm}
\item[(e).]{\it each column triangle reaches the maximal area with its
$e$-side fixed;}

\vspace*{-2mm}
\item[(f).]{\it each column triangle is an isosceles triangle.}\\
\vspace*{-4mm}
\end{itemize}

After setting up the above preliminaries, we are ready to study the
origin of CMCPV in flavor symmetry in the next section.

\vspace*{2mm}
\section{Origin of Constrained Maximal CP Violation}
\label{sec:3}
\label{sec:A-Theorem-for}
\vspace*{1mm}

In this Section, we will trace CMCPV to the ``residual symmetries'',
i.e., the subgroups of the original flavor symmetry group that remain intact
after the full group is broken.

Consider that the flavor symmetry group $\,G\,$ is broken down to two residual
symmetries $\,G_{\nu}^{}\,$ and $\,G_{\ell}^{}\,$ for neutrinos and charged leptons,
respectively. They are defined as follows,
\beqa
G~\to~
\begin{cases}
G_{\nu}^{}\!: & \{S\,|\,S^{T}M_{\nu}^{}S = M_{\nu}^{}\} ;
\\[1.5mm]
G_{\ell}^{}\!: &
\{T\,|\,T^{\dagger}M_{\ell}^{}M_{\ell}^{\dagger}T = M_{\ell}^{}M_{\ell}^{\dagger}\} ;
\hspace*{4mm}
\end{cases}
\label{eq:1210}
\eeqa
where $M_{\nu}^{}$ is the Majorana mass matrix of neutrinos, and
$\,M_{\ell}^{}M_{\ell}^{\dagger}$\,
is the effective mass matrix of left-handed charged leptons.
Thus, the mixing matrices $\,U_{\nu}^{}$\, and $\,U_{\ell}^{}$\, (which diagonalize
$M_{\nu}^{}$ and $M_{\ell}^{}M_{\ell}^{\dagger}$,\, respectively)
are directly determined by $\,S\,$ and $\,T\,$ \cite{Lam},
\beq
\ba{lcl}
U_{\nu}^{\dag}\hspace{0.3mm}S\hspace{0.3mm} U_{\nu}^{} &\!\!=\!\!& D_{S}^{} \,,
\\[1mm]
U_{\ell}^{\dag}\hspace{0.3mm}T\hspace{0.3mm} U_{\ell}^{} &\!\!=\!\!& D_{T}^{} \,.
\ea
\label{eq:1210-1}
\eeq
Here the matrices $D_S^{}$ and $D_T^{}$ are diagonal matrices.
Since Eq.\,(\ref{eq:1210-1}) demonstrates a direct connection between
flavor symmetry and lepton mixings, it also attracted extensive studies
\cite{flv-th,He:2011kn,He:2012yt,Rodejohann:2015pea} via the approach of
symmetry and group theory, without resorting to explicit mass matrices
or a fundamental Lagrangian.
With this {\it general mass-independent approach,} we will
analyze the origin of CMCPV.

Roughly speaking, our theorem states that the CMCPV can be realized
if the residual symmetries are real.
In rigorous manner, we formulate this theorem in the following form.
\\[2mm]
\noindent{\bf Theorem\,2}~[Origin of CMCPV].
\\[0.5mm]
{\it If the residual flavor symmetries in the lepton sector
(including charged leptons and neutrinos) are real
and fully determine the mixing pattern, then the CMCPV always holds, up to
a few minor exceptions:}
\begin{itemize}

\vspace*{-2mm}
\item[(i).]\,{\it one of the three mixing angles in the PMNS matrix is zero;}

\vspace*{-2mm}
\item[(ii).]\,{\it neutrinos are not Majorana fermions;}

\vspace*{-2mm}
\item[(iii).]\,{\it the residual symmetry for charged leptons is a Klein group,
i.e., $G_{\ell}^{} = Z_2^{}\otimes Z_2^{}$\,.}

\end{itemize}
It is clear that the exception\,(i) is already excluded by current oscillation data,
and the exception\,(ii) is not a concern for most neutrino theories.
The exception\,(iii) is less trivial, but can be easily evaded in model-buildings.
Besides, in Theorem\,2, for the residual symmetries being real, we mean that
there always exists a basis under which these symmetries become real.

To illustrate this theorem explicitly, we first consider a simple
(unrealistic) example.
A rotation of $120^{\circ}$ around the axis $(1,1,1)^{T}$ can be
represented by
\begin{equation}
R \,=\, \left(\!
\begin{array}{ccc}
0 & 1 & 0\\
0 & 0 & 1\\
1 & 0 & 0
\end{array}\!\right) .
\label{eq:0621-1}
\end{equation}
Suppose that this $\,R\,$ corresponds to the residual symmetry group $\,G_{\!\ell}^{}$\,.\,
Thus, we have
\beqa
U_{W}^{\dag}R\,U_{W}^{} \,=\, \text{diag}(1,\,\omega,\,\omega^{2})\,,
\eeqa
where $\,U_{W}^{}$ is the Wolfenstein mixing matrix defined in Eq.\,(\ref{eq:0621}).
According to the relation (\ref{eq:1210-1}), we have $\,U_{\ell}^{}=U_W^{}$,\,
which shows that $\,U_{\ell}^{\dag}$\, exhibits RCE.
If we further assume that the neutrino
mass matrix is diagonal (and thus $U_{\nu}^{}=I$\,),
then the PMNS matrix $\,U=U_{\ell}^{\dag}U_{\nu}^{}$\, gives,
$\,(|\delta|,\,\theta_{23}^{})=(\frac{\pi}{2},\,\frac{\pi}{4})$,\,
because Theorem\,1 states that the RCE always leads to CMCPV.

Two remarks are in order for this example. One is that both
residual symmetries are real. This is explicit for $\,G_{\ell}^{}$\,.\,
The form of $\,G_{\nu}^{}\,$ is,
$\,G_{\nu}\supset\{\text{diag}(1,1,-1),\text{diag}(1,-1,1)\}$,\,
since neutrino mass matrix is taken to be diagonal.
The other point is that $\,U_{\ell}^{\dag}\,$ exhibits RCE.

In general, the validity of Theorem\,2 (excluding its exceptions) implies
the following important points:
\begin{itemize}
\vspace*{-2mm}
\item[1).]\,real $G_{\nu}^{}$ leads to real $U_{\nu}^{}$,\,
which will be explicitly proven in \ref{sec:a1};

\vspace*{-2mm}
\item[2).]\,real $G_{\ell}^{}$ leads to RCE in $U_{\ell}^{\dag}$,\,
which will be explicitly proven in \ref{sec:a2};

\vspace*{-2mm}
\item[3).]\,if $\,U_{\ell}^{\dag}\,$ has RCE and $\,U_{\nu}\,$ is real,
then the PMNS matrix $\,U=U_{\ell}^{\dag}U_{\nu}^{}\,$ has RCE.
\end{itemize}
The three points above are combined to prove Theorem\,2.
The last point is just based on the Lemma given above Eq.\,(\ref{eq:0707-7}),
and the first two points can be understood
by the following reasoning. (We delegate the mathematical proofs
to \ref{App:B}.)  Since both $G_{\ell}^{}$ and $G_{\nu}^{}$ contain only
real transformations, they can be geometrically regarded as rotations
in 3-dimensional Euclidean space. (Here a trivial minus sign between the
determinants of SO(3) and O(3) does not matter.)
For such a rotation represented by a matrix $\,R$\,,\,
the rotation axis is one of its eigenvectors with the corresponding
eigenvalue equal to 1. The remaining two eigenvectors must be complex
conjugate to each other, which is a general property of \,SO(3)\, matrices.
(The explicit forms of the two eigenvectors are given in \ref{sec:a2}.)
Hence, if $\,R\in G_{\ell}^{}\,$,\, then the eigenvectors of $\,R\,$ are the columns
of $\,U_{\ell}^{}$\,,\, which implies two columns in $\,U_{\ell}^{}\,$ are conjugate
to each other, and thus $\,U_{\ell}^{\dag}$\, has RCE.

There are differences in the neutrino sector, because we consider
the neutrinos as Majorana particles here.
Hence, the residual symmetry has to be constructed with $Z_2^{}$'s,
i.e., $\,G_{\nu}^{}=Z_2^{}\otimes Z_2^{}\,$,\,
which geometrically correspond to two rotations of $180^{\circ}$.\,
These are special rotations in the sense that only such rotations may commute with
rotations around different axes.
For rotations of $180^{\circ}$,\, the eigenvalues are \,$(1,-1,-1)$,\,
cf.\ Eq.\,\eqref{eq:0708-6}.
Due to a partial degeneracy of the eigenvalues,
the neutrino mass matrix $\,M_{\nu}^{}\,$ should be determined
by two $Z_2^{}$-rotations with orthogonal axes. Each axis
determines one column of $\,U_{\nu}^{}$,\, so $\,U_{\nu}^{}\,$ only contains real
column vectors. This in turn implies that $\,G_{\ell}^{}\,$ cannot be
$\,Z_2^{}\otimes Z_2^{}$,\,
which is the exceptional case\,(iii)  pointed out in Theorem\,2:
if $\,G_{\ell}^{}=Z_2^{}\otimes Z_2^{}\,$,\,
then $\,U_{\ell}^{}\,$ will be real and no CP violation exists.
Now, it is also easy to understand why Theorem\,2 requires neutrinos
to be Majorana fermions, since the symmetries
$\,Z_2^{}\otimes Z_2^{}$\, are needed for $\,G_\nu^{}\,$.

Theorem\,2 further leads to a series of corollaries which we will discuss
as follows.
\\[2mm]
\noindent{\bf Corollary\,A}~$[$\,O(3) Subgroups\,$]$:
\\[0.5mm]
{\it If an O(3) subgroup $G$ contains sufficient residual symmetries
that can fully determine a mixing matrix, then it leads to the CMCPV after avoiding
the three exceptions listed in Theorem\,2.}
\\[-1mm]

This is manifest because the constraint which requires the residual symmetries
to be subgroups of O(3) makes $\,G_{\ell}^{}\,$ and $\,G_{\nu}^{}\,$
automatically real. According to Theorem\,2, this leads to the CMCPV.
Examples of such residual symmetries include popular groups like
$A_{4}^{}$, $S_{4}^{}$, and $A_{5}^{}$, corresponding
to tetrahedral, octahedral, and icosahedral symmetries, respectively.

We should comment on the phrase ``sufficient residual symmetries''
in Corollary\,A. As is well-known, the maximal residual symmetries
in the charged lepton and neutrino sectors are
$\,U(1)\otimes U(1)\otimes U(1)$\, and
$\,Z_2^{}\otimes Z_2^{}\otimes Z_2^{}$,\, respectively\,\cite{Lam}.
But when seeking flavor groups to unify the residual symmetries,
it is not necessary to cover those large groups.
For charged leptons, the minimal choice is to take a $\,Z_3^{}\,$ subgroup
from those $U(1)$'s, which is in fact sufficient to determine the mixing $\,U_\ell^{}$\,.\,
For the neutrino sector, the minimal sufficient residual symmetry should be
$\,Z_2^{}\otimes Z_2^{}$.\, So, this smaller set
of residual symmetries should be included in the O(3) subgroup
for the Corollary\,A.

However, in some models, especially those based on $A_4^{}$,\,
sometimes the flavor group does not contain sufficient residual symmetries,
and the so-called accidental symmetries are present to fully determine the mixings.
Those accidental symmetries depend on the detailed dynamics of the model
(instead of the flavor group), so the Corollary\,A does not apply.
But, if the accidental symmetry is a real symmetry, then Theorem\,2
applies and there is still CMCPV.
\\[-1mm]

\noindent{\bf Corollary B } [\,Real $M_{\nu}$\,]:
\\[0.5mm]
\emph{If $\,G_{\ell}^{}\,$ is real and $\,M_{\nu}^{}\,$ is real
or $\,M_{\nu}^{}\,$ can be written as
$\,M_{\nu}=z_{1}^{}{\cal I}+z_{2}^{}\tilde{M}_{\nu}^{}$,\,
where $\,\tilde{M}_{\nu}^{}$\, and $\,{\cal I}\,$ are real and identity matrices,
respectively, and $\,(z_1^{},\,z_2^{})$\,  are complex numbers, then there is CMCPV,
after evading the three exceptions listed in Theorem\,2.}
\\[-3mm]

We first consider the case that $M_{\nu}^{}$ is real. Then, as a real symmetric matrix,
$M_{\nu}^{}$ can be diagonalized by a real orthogonal matrix, which implies
$\,U_{\nu}^{}\,$ and $\,G_{\nu}^{}\,$ are real. Hence, according to Theorem\,2, we have
CMCPV. Multiplying $\,M_{\nu}^{}\,$ by an overall complex phase
will not change $\,U_{\nu}^{}\,$.\, Thus, if $\,M_{\nu}^{}=z_2^{}\tilde{M}_{\nu}^{}\,$
with real $\,\tilde{M}_{\nu}^{}$,\, then this means that $\,M_{\nu}^{}\,$ is essentially
real, up to an overall complex phase factor.  Hence, this case also leads
to CMCPV.  Next, consider
$\,M_{\nu}=z_{1}{\cal I}+\tilde{M}_{\nu}$\, with real $\,\tilde{M}_{\nu}^{}$.\,
This means that $\,M_{\nu}^{}\,$ is real up to subtracting a constant from all
diagonal elements. In this case, for $\,S\,(\in G_\nu^{})\,$ satisfying
$\,S^{T}\tilde{M}_{\nu}^{}S=\tilde{M}_{\nu}^{}$\,,\,  we have
$\,S^{T}M_{\nu}^{}S =S^{T}(z_1^{}{\cal I} + \tilde{M}_{\nu}^{})S
 = z_{1}^{}{\cal I} + \tilde{M}_{\nu}^{}$,\,
which shows that $\,M_{\nu}^{}\,$ is also invariant under $S$.\,
Hence, $\,M_{\nu}^{}\,$ has invariance under real $\,G_{\nu}^{}\,$
and there is CMCPV.
Finally, combining the two cases above, we have thus proven the Corollary\,B
for the general form $\,M_{\nu}^{}=z_{1}^{}{\cal I}+z_{2}^{}\tilde{M}_{\nu}^{}$\,.

The form $\,M_{\nu}^{}=z_{1}^{}{\cal I}+z_{2}^{}\tilde{M}_{\nu}^{}$\,
has important applications in model buildings for CMCPV.
Typically, for building flavor symmetry models, at least one
flavon $\,\phi\,$ is introduced to couple with neutrinos $\,\nu\,$
and forms a Yukawa term $\,\nu\nu\phi\,$,\,
which contributes to neutrino masses if the vacuum expectation values (VEV)
$\,\langle\phi\rangle\neq 0\,$,\,
where $\,\phi\,$ is a scalar field acting as a multiplet under the flavor symmetry.
Neutrinos are commonly considered as flavor triplets in many models,
so a $\,\nu\nu\,$ term or $\,\nu\nu\,\xi\,$ term
will usually show up, where $\,\xi\,$ is a flavor singlet.
These terms will contribute to $\,M_{\nu}^{}\,$ as a diagonal mass term
$\,z_1^{}{\cal I}\,$,\, where $\,z_1^{}\,$ is complex because the coefficients (Yukawa
couplings) of these terms are complex in general. The $\,\nu\nu\phi\,$
term will contribute in a form of $\,z_2^{}\tilde{M}_{\nu}^{}\,$
if $\,\langle\phi\rangle\,$  is real (up to an overall complex phase)
and the Clebsch-Gordan (CG) coefficients for the term are real.

Note that here we only consider the case with all 3 generations of neutrinos
unified into a triplet of the flavor group. Otherwise, the $\,\nu\nu\,$ term would not be
diagonal.\footnote{For certain groups, such as $\Delta(3n^2)$ or $\Sigma(3n^3)$
with $\,n\geqq 3\,$,\, two triplets cannot form a singlet. Those are certain subgroups of SU(3),
to be precise, subgroups with faithful irreducible 3-dimensional representation
whose determinant equals 1\,,\, that have complex representations and are not subgroups of SO(3).
In this case, the $\,\nu\nu\,$ term is absent, which means that in the general form
$\,M_{\nu}^{} = z_1^{}I + z_2^{}\tilde{M}_{\nu}^{}\,$
only $\,z_2^{}\tilde{M}_{\nu}^{}\,$ exists. So the problem becomes simpler.}\,
The flavon $\,\phi$\, can be in any non-trivial representation.
We should point out that both real $\,\langle\phi\rangle\,$ and real
CG coefficients are very common in many groups. For instance, in the
3-dimensional representation of $\,A_4^{}$,\, the CG coefficients for
$\,3\otimes 3\otimes 3\to 1$\, are real in both
the real basis (used for example in \cite{Ma:2001dn}) and the complex
basis (used for example in \cite{Altarelli:2005yx}).
This corollary does not apply to groups with inherent complex CG coefficients,
like $\,T'\,$ \cite{Tp} or $\,\Delta(27)\,$ \cite{delta27}.
As for real $\,\langle\phi\rangle$\,,\, if $\,\phi\,$ is a real
scalar field by definition,  then $\,\langle\phi\rangle\,$ is real.
If $\,\phi\,$ has to be a complex field, then as known from
minimization of scalar potentials in many models, it is still
common to have VEV alignment according to
$\,\langle\phi\rangle\propto (1,1,1),\, (1,0,0)$,\, etc., which is real.
If $\langle\phi\rangle$ is however complex, then in general CMCPV does not follow.
For those ``real $\nu\nu\phi$ models'', where only the $\,\nu\nu\phi\,$ term makes
a non-trivial contribution (not proportional to the unit matrix) to $\,M_\nu^{}$,\,
we have the following corollary.
\\[2mm]
\noindent{\bf Corollary\,C}~[\,Real $\nu\nu\phi$ Models\,]:
\\[0.5mm]
{\it Real $\nu\nu\phi$ models always lead to CMCPV\,,\, if the three minor exceptions
listed in Theorem\,2 do not happen.}
\\[-3mm]

For a demonstration of the above general discussion,
we will build a simple ``real $\nu\nu\phi$ model''
in the following Section\,\ref{sec:nnp}.

\vspace*{1mm}
\section{Applications }
\label{sec:Application}
\label{sec:4}

In this section, we apply our theorems and corollaries to various situations
and understand why CMCPV is realized in certain cases.
We will illustrate how to achieve the CMCPV in model buildings.
There are extensive recent literature \cite{mutau} studying specific models of
$\,(|\delta |,\,\theta_{23}^{}) =(\frac{\pi}{2},\,\frac{\pi}{4})\,$.

\subsection{$\mu -\tau$\, Reflection Symmetry}
\label{sub:reflection}
\label{sub:4.1}
\vspace*{2mm}

The $\mu$-$\tau$ reflection symmetry was studied before
\cite{Harrison:2002et,Ma:2002ce,Grimus:2003yn,MR}, which is
sometimes also called the generalized $\mu$-$\tau$ symmetry.
This symmetry is defined in the flavor eigenbasis
(with $\,M_{\ell}^{}\,$ diagonal),
\begin{equation}
\nu_{e}^{}    \rightarrow\nu_{e}^{*},~~~~
\nu_{\mu}^{}  \rightarrow\nu_{\tau}^{*},~~~~
\nu_{\tau}^{} \rightarrow\nu_{\mu}^{*}\,.
\label{eq:0702-1}
\end{equation}
Imposing this symmetry leads to the following form of the neutrino and lepton mass matrices,
\begin{equation}
\tilde{M}_{\nu}^{} =  \left(\!\begin{array}{ccc}
r_{1}^{} & z_{1}^{} & z_{1}^{*}
\\
. & z_{2}^{} & r_{2}^{}
\\
. & . & z_{2}^{*}
\end{array} \!\right) ,
\hspace*{10mm}
\tilde{M}_{\ell}^{2} =  \left(\!
\begin{array}{ccc}
m_{e}^{2} & &    \\
 & m_{\mu}^{2} & \\
 &  & m_{\tau}^{2}
\end{array} \!\right) ,
\label{eq:0520}
\end{equation}
where $\,\tilde{M}_{\nu}^{}\,$ is symmetric and
$\,\tilde{M}_{\ell}^2 \equiv\tilde{M}_{\ell}^{}\tilde{M}_{\ell}^{\dag}\,$ is diagonal.
Note that the elements $\,r_{1,2}^{}\,$ are real, but $\,z_{1,2}^{}\,$
are complex in general.  The operation (\ref{eq:0702-1}) will transform
$\,\nu^{T}\tilde{M}_{\nu}\,\nu$\,
to its Hermitian conjugate.  We can directly check that the Lagrangian term,
$\,\mathcal{L}\supset\nu^{T}\tilde{M}_{\nu}^{}\nu+\text{h.c.}$,\,
is invariant under the transformation (\ref{eq:0702-1})
if and only if $\,\tilde{M}_{\nu}^{}\,$ takes the form of Eq.\,(\ref{eq:0520}).

Let us make a transformation,
\begin{equation}
M_{\nu}^{} \,=\, U_{\ell}^{}\tilde{M}_{\nu}^{}U_{\ell}^{T},
\hspace*{7mm}
M_{\ell}^{2} \,=\, U_{\ell}^{}\tilde{M}_{\ell}^{2}U_{\ell}^{\dagger}\,,
\label{eq:0702}
\end{equation}
with
\begin{equation}
U_{\ell}^{} \,=\, \left(\!
\begin{array}{ccc}
1 & 0 & 0
\\[0.5mm]
0 & \frac{1}{\sqrt{2}\,} & \frac{1}{\sqrt{2}\,}
\\[1.5mm]
0 & -\frac{\ii}{\sqrt{2}\,} & \frac{\ii}{\sqrt{2}\,}
\end{array}\!\!\right) .
\label{eq:0520-2}
\end{equation}
Thus, we derive
\begin{equation}
M_{\nu}^{} =\, \left(\!
\begin{array}{ccc}
r_1^{} & \sqrt{2}\,z_{11}^{} & \sqrt{2}\,z_{12}^{}
\\[0.5mm]
. & r_2^{}\!+\!z_{21}^{} & z_{22}^{}
\\[0.5mm]
. & . & r_2^{}\!-\!z_{21}^{}
\end{array}\!\right) ,
\hspace*{10mm}
M_{\ell}^{2} =\, \left(
\!\begin{array}{ccc}
a & 0 & 0  \\[0.5mm]
0 & ~b_+^{} & \ii\hspace*{0.3mm} b_-^{} \\[0.5mm]
0 & -\ii\hspace*{0.3mm} b_-^{} & ~b_+^{}
\end{array}\!\right) ,
\hspace*{5mm}
\label{eq:0520-1}
\end{equation}
where we have defined notations,
$\,z_j^{}\equiv z_{j1}^{}+\ii z_{j2}^{}\,$,\, ($j=1,2$),
and $\,a\equiv m_{e}^{2}$\,,\, $b_\pm^{} \equiv \frac{1}{2}(m_{\mu}^{2}\pm m_{\tau}^{2})\,$.\,
The quantities $(z_{j1}^{},\,z_{j2}^{})$ and $(a,\,b_\pm^{})$ are all real.
Note that $\,M_{\nu}^{}\,$ is a real matrix, and
the charged lepton sector has an SO(2) residual symmetry,
\begin{equation}
R \,=\, \left(\begin{array}{ccc}
1 & 0 & 0 \\[0.4mm]
0 & \cos\theta & \sin\theta \\[0.4mm]
0 & -\sin\theta & \cos\theta
\end{array}\right) .
\label{eq:0703}
\end{equation}
This is because $\,RM_{\ell}^{2}R^{\dagger}=M_{\ell}^{2}\,$ holds for
$\,\theta \in [0,\,2\pi )$\,.\, Since $\,M_{\nu}^{}\,$ and $\,G_{\ell}^{}\,$ are all real,
this will lead to CMCPV according to our Corollary\,B.
Note that real $\,M_{\nu}^{}\,$ implies that $\,G_{\nu}\,$ is real.

The $\mu$-$\tau$ reflection symmetry is certainly not the
only possibility to generate CMCPV.  From Eq.\,(\ref{eq:0520-1}),
we see that $\,M_{\nu}^{}\,$ is real, while Corollary\,B shows that $\,M_{\nu}^{}\,$
can have a more general form including complex numbers.
Hence,  the $\mu$-$\tau$ reflection symmetry is just a special case
of real residual symmetries, although this is not manifest before the transformation
of basis in Eq.\,(\ref{eq:0702}).

\vspace*{2mm}
\subsection{CMCPV from Geometrical Symmetry Breaking}
\label{sub:4.2}
\vspace*{2mm}

As another example illustrating our theorem,
we revisit a model from Ref.\,\cite{He:2012yt}, which predicted
$\,|\delta| = \frac{\pi}{2}$\, and $\,\theta_{23}^{} = \frac{\pi}{4}$\,
(as well as $\,\theta_{13}^{}\simeq \frac{\pi}{4}-\theta_{12}^{}$).
We will show that this model fulfills the criteria for CMCPV.

This model identifies a $\,Z_{4}^{}\,$ rotation around the $x$-axis as
$\,G_{\ell}^{}\,$,\, and the product reflections $\,Z_2^{}\otimes Z_2^{}$\,
as $\,G_{\nu}^{}$,\, where one $\,Z_2^{}\,$ reflects $\,y\to -y\,$ and
the other $\,Z_2^{}\,$ transforms $\,(x,\,z)\to -(z,\,x)\,$.\,
These rotations are subgroups of the octahedral symmetry $\,O_h^{}$\,,\,
and can be shown by simple geometrical picture.
This group setting generates the bimaximal mixing,
$\,\theta_{12}^{}=\theta_{23}=\frac{\pi}{4}$\,
and $\,\theta_{13}^{}=0$.\,
The necessary deviation from this leading order scheme
was generated by slightly tilting the axis of $\,Z_4^{}\,$ rotation
by a small angle (defined as $\!\sqrt{2}\epsilon$\,)
that turns out to be related to nonzero $\,\theta_{13}^{}$\,.\,
We also verified that this geometrical symmetry breaking
can arise from certain flavon models.
For example, we may set up a concrete realization,
where a flavor triplet $\,\phi\,$ is responsible for mass-generation of
the charged leptons and the Yukawa terms involving $\,\phi\,$ are \,SO(3)\, symmetric
in the 3-dimensional flavor space. With these, the geometrical breaking is
connected to the VEV misalignment of flavons. After the axis tilt, the residual
symmetry of charged lepton mass matrix is represented by \cite{He:2012yt},
\begin{equation}
R_{\ell}^{} \,=\, \left(\!\begin{array}{ccc}
1 & -2\epsilon & 0\\
0 & 0 & -1\\
2\epsilon & 1 & 0
\end{array}\!\right) + {\cal O}(\epsilon^{2})\,.
\label{eq:0708-8}
\end{equation}
The neutrino mass matrix is still invariant under the original reflections
$\,Z_2^{}\otimes Z_2^{}$,\, as represented by \cite{He:2012yt},
\begin{equation}
R_{\nu 1}^{} =\, \left(\!\!\begin{array}{ccc}
0 & 0 & -1\\
0 & 1 & 0\\
-1 & 0 & 0
\end{array}\!\right) ,
\hspace*{8mm}
R_{\nu 2}^{} =\, \left(\!\begin{array}{ccc}
1 & 0 & 0\\
0 & -1 & 0\\
0 & 0 & 1
\end{array}\!\right) .
\label{eq:0708-9}
\end{equation}
Since all the residual symmetries are real, Theorem\,2 applies and
the model should realize CMCPV.  This is indeed the case,
as found in Ref.\,\cite{He:2012yt}.

\vspace*{2mm}
\subsection{A Real $\nu\nu\phi$ Model}
\label{sec:nnp}
\label{sec:4.3}
\vspace*{2mm}

As stated in Corollary\,C, the real $\,\nu\nu\phi\,$ models should always produce
CMCPV.\, In the following, we will build such a model as an explicit illustration.

We use $\,A_4^{}\otimes Z_2^{}\,$ as flavor symmetry group and introduce
two scalar fields $\,\phi^{\ell}\,$ and $\,\phi^{\nu}$,\, in addition to
the SM Higgs doublet $H$. The relevant particle content of this model is summarized
in Table\,\ref{tab:1}. The Lagrangian for the lepton-neutrino sector contains,
\begin{eqnarray}
\mathcal{L} & \,\supset\, & y_{e}^{}(L\phi^{\ell})He^{c}
+ y_{\mu}^{}(L\phi^{\ell})'H\mu^{c} + y_{\tau}^{}(L\phi^{\ell})''H\tau^{c}
\nonumber\\[1mm]
 & & +\,y_{\nu1}^{}(LL)HH + y_{\nu2}^{}(\phi^{\nu}LL)HH+\rm{h.c.},
\label{eq:0707}
\end{eqnarray}
where $\,L\,$ stands for the left-handed lepton doublet of SU(2)$_L^{}$
and $\,H\,$ is the Higgs doublet.
Since $\,\phi^{\ell}\,$ and $\,\phi^{\nu}\,$ do not have any charge other than
the $Z_2^{}$ assignment, they can be
real fields.
Consider that they acquire the following alignment of VEVs,
\beqa
\langle\phi^{\nu}\rangle\propto (1,\,\epsilon_2^{},\,\epsilon_3^{})\,,
\hspace*{5mm}
\langle \phi^{\ell} \rangle \propto (1,1,1)\,,
\eeqa
where similar to \cite{Barry:2010zk},
we have introduced a small perturbation on the usual VEV alignment in
$\,\langle\phi^{\nu}\rangle$\,,\, with $\,\epsilon_2^{},\epsilon_3^{}\ll 1\,$.\,
Similar VEV alignment was already considered in the literature\,\cite{Barry:2010zk},
but its further elaboration 
is irrelevant to the current illustration purpose of realizing CMCPV
on the ground of residual symmetries\,\cite{Lam};
it is also fully beyond the main goal of this short Letter.
For $\,\epsilon_2^{}=\epsilon_3^{}=0$\,,\, one would obtain the tri-bimaximal mixing;
and the small $(\epsilon_2^{},\,\epsilon_3^{})$
should produce the necessary corrections \cite{Barry:2010zk}.
Note that for $\,\epsilon_2^{}=\epsilon_3^{}=0$\,,\,
although  $\,\langle\phi^{\nu}\rangle$\, is real
as required by our definition of real $\nu\nu\phi$ models,
one of the mixing angles, $\theta_{13}^{}$, is zero,
which just matches the exception-(i) of our Theorem\,2.
Hence, CMCPV does not follow in this case.
For $\,\epsilon_{2,3}^{}\neq 0\,$,\, the charged leptons and neutrinos acquire
masses as follows,
\begin{equation}
M_{\ell}^{} \,\propto\, \left(\!
\begin{array}{cll}
a~ & b & c \\[0.5mm]
a~ & b\hspace*{0.3mm}\omega^{2} & c\hspace*{0.3mm}\omega
\\[0.5mm]
a~ & b\hspace*{0.3mm}\omega & c\hspace*{0.3mm}\omega^{2}
\end{array}\!\!\right) ,
\hspace*{8mm}
M_{\nu}^{} \,\propto\, \left(\!\!\begin{array}{ccc}
d & \epsilon_{3}^{} & \epsilon_{2}^{} \\[0.5mm]
\epsilon_{3}^{} & d & 1 \\[0.5mm]
\epsilon_{2}^{} & 1 & d
\end{array}\!\!\right) ,~~~
\label{eq:0707-2}
\end{equation}
where the mass parameters \,$(a,\,b,\,c,\,d)$\, are complex in general.
This type of lepton and neutrino mass matrices are often studied
in the literature\,\cite{abcd}. For instance, diagonalizing the lepton mass
matrix $\,M_\ell^{\dag}M_\ell^{}\,$ gives the mass-eigenvalues
$\,(m_e^{},\,m_\mu^{},\,m_\tau^{}) \propto (|a|,\,|b|,\,|c|)\,$.\,
[This also shows that using the observed mass values $\,(m_e^{},\,m_\mu^{},\,m_\tau^{})\,$
does not fully fix the parameters \,$(a,\,b,\,c)$\, themselves; while inputting the
model parameters \,$(a,\,b,\,c)$\, can fully accommodate the observed lepton masses.]
The focus of our paper is on the origin of leptonic mixings from flavor symmetry.
Thus, by diagonalizing $\,M_\ell^{}\,$ and $\,M_\nu^{}\,$,\,
we derive the following lepton and neutrino mixing matrices,
\beqa
U_{\nu}^{} \,\simeq\,
\left(\!\!\begin{array}{ccc}
\frac{\epsilon_{2}^{}+\epsilon_{3}^{}}{\sqrt{2}}
& 1 & \frac{-\epsilon_2^{}+\epsilon_3^{}}{\sqrt{2}}
\\[2mm]
\frac{1}{\sqrt{2}\,} & -\epsilon_2^{} & -\frac{1}{\sqrt{2}\,}
\\[2mm]
\frac{1}{\sqrt{2}\,} & -\epsilon_3^{} & \frac{1}{\sqrt{2}\,}
\end{array}\!\!\right) ,
\label{eq:0707-3}
\eeqa
\vspace*{-3mm}
and
\beqa
U_{\ell}^{} \,=\, \frac{1}{\sqrt{3}\,}
\left(\!\begin{array}{cll}
1 & \,1 & \,1\\
1 & \omega^{2} & \omega\\
1 & \omega & \omega^{2}
\end{array}\!\!\right) .
\label{eq:0707-4}
\eeqa
From $\,U=U_{\ell}^{\dagger}U_{\nu}^{}$\,,\, it is straightforward to extract
the PMNS parameters in the standard parametrization,
\beqs
\beqa
&&
\theta_{23}^{}=\frac{\pi}{4}\,,
\hspace*{7mm}
|\delta|=\frac{\pi}{2}\,,
\label{eq:0707-5}
\\
&&
\sin{\theta_{13}^{}} \simeq \frac{\,\epsilon_3^{}\!-\! \epsilon_2^{}\,}{\sqrt{6}}\,,
\hspace*{7mm}
\tan\theta_{12}^{}\simeq
\frac{\,\sqrt{2}(1\!-\!\epsilon_2^{}\!-\!\epsilon_3^{})\,}
     {~2\!+\!\epsilon_2^{}\!+\!\epsilon_3^{}~}\,.
\hspace*{10mm}
\label{eq:0707-6}
\eeqa
\eeqs
These results show that,
apart from model-specific deviations of $\,(\theta_{13}^{},\,\theta_{12}^{})\,$
from their tri-bimaximal values, we have realized the CMCPV, as expected from
Corollary\,C.
From Eq.\,\eqref{eq:0707-6}, we can determine the perturbative parameters
$\,(\epsilon_{2}^{},\,\epsilon_{3}^{})\,$ in terms of
$\,(\sin\theta_{13}^{},\,\tan\theta_{12}^{})\,$ via
\begin{equation}
\epsilon_{2}^{} \,=
-\sqrt{\frac{3}{2}}\sin\!\theta_{13}^{}
+\frac{\,1\!-\!\!\sqrt{2}\tan\!\theta_{12}^{}\,}{\,2\!+\!\!\sqrt{2}\tan\!\theta_{12}^{}\,}\,,
\hspace*{8mm}
\epsilon_{3}^{} \,=
\sqrt{\frac{3}{2}}\sin\!\theta_{13}^{}
+\frac{\,1\!-\!\!\sqrt{2}\tan\!\theta_{12}^{}\,}{\,2\!+\!\!\sqrt{2}\tan\!\theta_{12}^{}\,}\,.
\label{eq:0806}
\end{equation}
Taking $\theta_{13}^{}\simeq 9^{\circ}$ and $\theta_{12}^{}\simeq 34^{\circ}$,\,
we derive $\,(\epsilon_2^{},\epsilon_{3}^{})\simeq (-0.18,\,0.21)$\,.\,

\begin{table}[t]
\centering
\protect\caption{Particle content of the $\,A_4^{}\otimes Z_2^{}\,$ model.}
\vspace*{2mm}
\begin{tabular}{c||c|c|c|c|c}
\hline
\hline
~Groups~ & $L$ & $~(e^{c},\,\mu^{c},\,\tau^{c})$~
 & $~\phi^{\ell}~$ & $~\phi^{\nu}~$ & ~$H$~~
\tabularnewline
\hline
$~A_4^{}$ & $3$ & $(1,\,1'',\,1')$ & $3$ & $3$ & $1$~
\tabularnewline

$~Z_2^{}$ & $\,-1~$~ & $1$ & $-1$~~ & $1$ & $1$~
\tabularnewline

~SU(2)$_L^{}$ & $2$ & $1$ & $1$ & $1$ & $2$~
\tabularnewline
\hline
\hline
\end{tabular}
\label{tab:Particle}
\label{tab:1}
\end{table}

\vspace*{1mm}

In general, we can extend the real $\nu\nu\phi$ models to type-I neutrino seesaw.
In this case, we may introduce three right-handed neutrinos $\,\nu_R^{}\,$ in the 3-dimensional
representation of $A_4^{}$.\, Thus, the neutrino Dirac mass matrix will be proportional
to unit matrix, $\,m_D^{}\propto {\cal I}\,$,\,
while the heavy Majorana mass matrix $\,M_R^{}\,$ shares similar structure
with $\,M_\nu^{}\,$ in Eq.\,(\ref{eq:0707-2}).
Hence, we find that the seesaw mass matrix of light neutrinos
$\,M_\nu^{} \propto M_R^{-1}\,$.\,

\section{Conclusions}
\label{sec:con}
\label{sec:5}

In this work, we stressed that
a general parametrization-independent definition of the maximal CP violation should
be constructed in terms of Jarlskog invariant $\,J$\,,\, rather than the CP angle $\,\delta\,$
(which is rephasing non-invariant). We pointed out that naively maximizing $\,J\,$
without constraint is already excluded by oscillation data.
We further demonstrated the crucial importance of introducing the new concept of
constrained maximal CP violation (CMCPV) for studying the viable maximal CP violation.
For this purpose, we constructed CMCPV in the Definition\,1, and formulated it
by a set of equivalent ways, as summarized in Theorem\,1 (Sec.\,2).
We derived the parametrization-independent realization of
the CMCPV via solution \eqref{eq:|U2j|=|U3j|}, which was proven to be
the maximum of Jarlskog invariant
under a minimal constraint on the PMNS matrix $U$ (Sec.\,2 and Appendix\,A).
We found that the CMCPV just corresponds to
$\,(|\delta|,\,\theta_{23}^{})=(\frac{\pi}{2},\,\frac{\pi}{4})\,$
in the standard parametrization of the PMNS matrix \eqref{eq:0623}.
In Sec.\,3 and Appendix\,B, we proved Theorem\,2, stating
that if the residual symmetries of neutrinos and charged leptons
are real, then the CMCPV should be realized, up to a few minor exceptions.
It was shown that the conditions for CMCPV are actually quite common, and
we presented several sample models in Sec.\,4, demonstrating that in particular the often
considered $\mu$-$\tau$ reflection symmetry is a special case of our theorem.
We also note that the current formulation cannot be naively applied to the quark sector.
The reason is that our Theorem\,1 proves RCE to be essential for the CMCPV,
but RCE cannot hold for the CKM mixing matrix due to experimental data.
Namely, any two rows (or columns) in the CKM matrix cannot be conjugate to each other
(up to rephasing).

\vspace*{1mm}

If indeed the values of $\,\delta \simeq -\frac{\pi}{2}\,$ and
$\,\theta_{23}^{} \simeq \frac{\pi}{4}\,$ continue to be favored by neutrino data,
our general theorems and corollaries of CMCPV should be important, and provide strong
guidelines for the model buildings with flavor symmetry.

\vspace*{2mm}
\begin{appendix}

\section{Parametrization-Independent Solution of CMCPV}
\label{App:A}

In this Appendix, we derive the general solution of CMCPV by using the manifestly
parametrization-independent formula of Jarlskog invariant \eqref{eq:J2-all}.

Following our Definition\,1 for CMCPV, we can use Eq.\,\eqref{eq:J2-emu13}
to derive the extremal conditions of Jarlskog invariant respect to
$|U_{\mu 1}|$ and $|U_{\mu 3}|$ by fixing $|U_{e1}|$ and $|U_{e3}|$.
Thus, we have
\beqs
\beqa
\label{eq:dJ2/dz}
\frac{\dif J^2}{\dif\, z} &\!\!=\!\!&
xyw - \frac{1}{2}(1\!-y)\left[(1\!-y)\,z +(1\!-x)\,w -(1\!-x-y)\right]
\,=\, 0  \,,
\\[1.5mm]
\label{eq:dJ2/dw}
\frac{\dif J^2}{\dif\, w} &\!\!=\!\!&
xyz - \frac{1}{2}(1\!-x)\left[(1\!-y)\,z +(1\!-x)\,w -(1\!-x-y)\right]
\,=\, 0 \,,
\eeqa
\eeqs
where for convenience we have used the notations,
$\,(x,\,y,\,z,\,w)\equiv
(|U_{e1}|^2,\,|U_{e3}|^2,\,|U_{\mu 1}|^2,\,|U_{\mu 3}|^2)\,$.\,
From the extremal conditions \eqref{eq:dJ2/dz}-\eqref{eq:dJ2/dw}, we deduce
the solutions,
\beqa
\label{eq:zw}
z=\frac{1}{2}(1-x)\,, &&
w=\frac{1}{2}(1-y)\,.
\eeqa
Hence, we have
\beqa
\left|U_{\mu j}^{}\right|^2  \,=\,
\frac{1}{2}\(1-|U_{ej}^{}|^2\)
, \hspace*{8mm} (j=1,2,3)\,,
\label{eq:|U2j|}
\eeqa
where we have used the unitarity condition for the second row,
$\,\dis\sum_{j=1}^3 |U_{\mu j}|^2=1\,$.\,
With Eq.\,\eqref{eq:|U2j|} and making use of the unitarity conditions
for each column of the mixing matrix $\,U\,$,\, we further deduce
\beqa
\left|U_{\tau j}^{}\right|^2  \,=\,
\frac{1}{2}\(1-|U_{ej}^{}|^2\)
, \hspace*{8mm} (j=1,2,3)\,.
\label{eq:|U3j|}
\eeqa
Finally, comparing Eqs.\,\eqref{eq:|U2j|} abd \eqref{eq:|U3j|},
we arrive at
\beqa
\left|U_{\mu j}^{}\right|^2  \,=\,
\left|U_{\tau j}^{}\right|^2 \,=\,
\frac{1}{2}\(1-|U_{ej}^{}|^2\)
, \hspace*{8mm} (j=1,2,3)\,.
\label{eq:|U2j|=|U3j|-2}
\eeqa
This just reproduces the Eq.\,\eqref{eq:|U2j|=|U3j|},
which we presented in the text.

Next, we prove that the above extremal solution \eqref{eq:zw} or
\eqref{eq:|U2j|} indeed corresponds to a maximum of Jarlskog invariant.
For this purpose, we compute the second derivatives of the squared
Jarlskog invariant respect to $\,\(z,\,w\)$\,,
\beqa
\label{eq:J2"}
(J^2)''_{zz} \,=\,-\frac{1}{2}(1\!-y)^2,
\hspace*{5mm}
(J^2)''_{ww} \,=\,-\frac{1}{2}(1\!-x)^2,
\hspace*{5mm}
(J^2)''_{zw} \,=\, (J^2)''_{wz} \,=\,
-\frac{1}{2}(1\!-x-y-xy) \,.
\eeqa
Then, we inspect the eigenvalues
of the $2\!\times\! 2$ matrix $\,\{(J^2)''\}\,$,\,
whose elements are given by Eq.\,\eqref{eq:J2"}.
The eigenvalues $\,\{\lambda_1^{},\,\lambda_2^{}\}$\, satisfy the following
quadratic eigenvalue equation,
\beqs
\beqa
&& \lambda^2 - B\lambda + C \,=\, 0\,,
\\[1mm]
&&
B \,=\, (J^2)''_{zz} + (J^2)''_{ww}
=\, -\frac{1}{2}\left[(1\!-x)^2 \!+(1\!-y)^2\right] \,<\, 0
\,,
\\[1mm]
&&
C \,=\, (J^2)''_{zz}(J^2)''_{ww}\! -[(J^2)''_{zw}]^2
=\, xy\,(1\!-x-y) \,> 0\,,
\eeqa
\eeqs
where we have,
$\,1\!-x-y = 1\! - |U_{e1}|^2 - |U_{e3}|^2 = |U_{e2}|^2>0\,$,\,
due to the unitarity condition on the first row.
Thus, we have the two eigenvalues obey
$\,\lambda_1^{}\!+\lambda_2^{}=B <0\,$ and
$\,\lambda_1^{}\lambda_2^{} =C >0\,$.\,
This means that the two eigenvalues of $\,\{(J^2)''\}\,$
are both negative, $\,\lambda_1^{},\lambda_2^{} <0\,$.\,
Hence, we conclude that the extremal solution \eqref{eq:zw} or
\eqref{eq:|U2j|} is indeed the maximum of the Jarlskog invariant
(under the constraint on the first row of $U$),
and provides the parametrization-independent realization of the CMCPV
as given in our Definition\,1.

Finally, using the parametrization-independent solution
\eqref{eq:|U2j|=|U3j|-2} or \eqref{eq:|U2j|=|U3j|}
of CMCPV, we can readily derive the explicit realization of CMCPV
under the standard parametrization \eqref{eq:0623}.
From the first equality of Eq.\,\eqref{eq:|U2j|=|U3j|-2},
we have two independent conditions $\,|U_{\mu 1}|=|U_{\tau 1}|\,$
and $\,|U_{\mu 3}|=|U_{\tau 3}|\,$,\, which take the following forms
under the standard parametrization \eqref{eq:0623},
\beqs
\beqa
\label{eq:C1}
\left|s_{12}^{}c_{23}^{} \!+ c_{12}^{}s_{23}^{}s_{13}^{}e^{\ii\delta}\right|
&\hspace*{-2mm}=\hspace*{-2mm}&
\left|s_{12}^{}s_{23}^{}\! - c_{12}^{}c_{23}^{}s_{13}^{}e^{\ii\delta}\right| ,
\\[1mm]
\label{eq:C2}
\left|s_{23}^{}c_{13}^{}\right|
&\hspace*{-2mm}=\hspace*{-2mm}&
\left|c_{23}^{}c_{13}^{}\right|.
\eeqa
\eeqs
The condition \eqref{eq:C2} leads to $\,s_{23}^{}=c_{23}^{}\,$ and thus
$\,\theta_{23}^{}=\frac{\pi}{4}\,$.\,
Given this, we can rewrite \eqref{eq:C1} as
\beqa
\left|s_{12}^{} \!+ c_{12}^{}s_{13}^{}e^{\ii\delta}\right|
\,=\,
\left|s_{12}^{}\! - c_{12}^{}s_{13}^{}e^{\ii\delta}\right| .
\eeqa
Since $\,c_{12}^{}s_{13}^{}\neq 0\,$,
this must require $\,\cos\delta =0\,$,\, i.e.,
$\,|\delta|=\frac{\pi}{2}\,$.\,
Hence, the explicit realization of our CMCPV under the
standard parametrization \eqref{eq:0623} just gives
$\,(|\delta|,\,\theta_{23}^{})=(\frac{\pi}{2},\,\frac{\pi}{4})\,$,\,
which we mentioned in the text above Eq.\,\eqref{eq:0623}.

\section{Proofs}
\label{App:B}

In this Appendix, we present proofs that are needed to establish the Lemma given after
Eq.\,(\ref{eq:0623-3}) and the main Theorem\,2 given in Sec.\,\ref{sec:3}.

\vspace*{1mm}
\subsection{RCE is Invariant under Right-handed Real Transformations}
\label{sec:a3}
\vspace*{2mm}

For a unitary matrix $V$ with the form of ``row conjugation equality'' (RCE)
and a real orthogonal matrix $\,R$\,,\,
we need to prove that $\,V'=VR\,$ still has RCE. The proof is straightforward.
Defining the elements of these matrices,
\begin{equation}
V= (u_{ij}^{})\,,    \hspace*{4mm}
R= (r_{ij}^{})\,,    \hspace*{4mm}
V' \!= (u'_{ij})\,,  \hspace*{4mm}
\label{eq:0708-2}
\end{equation}
we have
\beqa
u'_{ij} \,= \sum_{k}u_{ik}^{}r_{kj}^{} \,.
\label{eq:0708-3}
\eeqa
Note that the matrix elements $\,(u_{1k}^{})\,$ and $\,(r_{kj}^{})$\,
(with $k,j=1,2,3$)\, are
real numbers from the start. The RCE feature of matrix $\,V\,$ gives,
$\,(u_{2j}^{})^{*}=u_{3j}^{}$\, for $j=1,2,3$.\,
This implies
\begin{equation}
u'_{1j}=\sum_{k}u_{1k}^{}r_{kj}^{}
= \textrm{real~numbers}\,,
\label{eq:0708-4}
\end{equation}
and
\beqa
(u'_{2j})^{*}=\sum_{k}u_{2k}^{*}r_{kj}=\sum_{k}u_{3k}r_{kj}=u'_{3j}\,.\label{eq:0708-5}
\eeqa
We have thus proven explicitly that RCE is invariant under right-handed real
transformations.

\vspace*{2mm}
\subsection{Real $G_{\nu}^{}$ Leads to Real $U_{\nu}^{}$}
\label{sec:a1}
\vspace*{2mm}

Consider Majorana neutrinos with residual symmetry
$\,G_{\nu}^{}=Z_2^{}\otimes Z_2^{}$.\,
In the following, we will prove that a real $\,G_{\nu}^{}\,$
leads to real $\,U_{\nu}^{}\,$,\, and vice versa.

Let us set $\,S\,$ to be a $3\!\times\!3$ unitary matrix, which is real and is a $\,Z_2^{}\,$
transformation (i.e., $S^{2}={\cal I}$\,).
As $\,S\,$ is real, it follows that $\,S^{\dag}=S^{T}\,$,\,
and the unitarity condition $\,SS^{\dag}={\cal I}\,$
implies that the real matrix $\,S\,$ is orthogonal,
$\,SS^{T}={\cal I}\,$.\, Without losing generality, we set $\,S\in \text{SO(3)}$.\,
Hence, $\,S\,$ is a rotation in 3-dimensional Euclidean space.
Furthermore, since $\,S^{2}={\cal I}$\,,\, it must be a $180^{\circ}$-rotation.

For $\,G_{\nu}^{}=Z_2^{}\otimes Z_2^{}$,\,  we may use $\,S_1^{}\,$ and $\,S_2^{}\,$
to represent the transformations of the two $Z_2^{}$'s, respectively.
Thus, $[S_1^{},\,S_2^{}]=0$\, should hold, which implies that their rotation axes must be
orthogonal.  Hence, geometrically $\,G_{\nu}^{}\,$ contains two $180^{\circ}$-rotations
with orthogonal axes. These two axes can be represented by two
normalized real vectors $\,v_1^{}$\, and $\,v_2^{}$\, with
%
\begin{equation}
\begin{array}{ll}
S_{1}^{}v_{1}^{} = v_{1}^{}\,,  ~~&~~ S_{1}^{}v_{2}^{} = -v_{2}^{}\,,
\\[1mm]
S_{2}^{}v_{1}^{} = -v_{1}^{}\,, ~~&~~ S_{2}^{}v_{2}^{}=v_{2}^{}\,,
\\[1mm]
\end{array}
\label{eq:0629}
\end{equation}
where $\,v_1^{}\,$ and $\,v_2^{}$ are column vectors, of the $3\!\times\!1$ matrix form.
Taking $\,v_3^{}=v_1^{} \!\times\! v_2^{}$\, and
$\,U_{\nu}^{} = (v_1^{},\,v_2^{},\,v_3^{})$,\,
we see that $\,U_{\nu}^{}\,$ is a real matrix and can diagonalize
$\,S_1^{}\,$ and $\,S_2^{}\,$ simultaneously in the way given by
Eq.\,(\ref{eq:1210-1}).

Therefore, if $\,G_{\nu}^{}\supset\{S_1^{},\,S_2^{}\}$\, contains only real matrices,
then $\,U_{\nu}^{}\,$ is real. The converse proposition that a real $\,U_{\nu}^{}\,$
leads to real $\,G_{\nu}^{}\,$ is also true, and can be readily proven.

\subsection{\label{sec:a2}Real $G_{\ell}^{}$ Leads to Complex $U_{\ell}^{\dag}$
with RCE}

We need to prove that any SO(3)\, matrix $\,R\,$ can be
diagonalized by $\,U_{R}^{\dag}R\,U_R^{}$\,,\, where the unitary matrix
$U_R^{}$ contains one real column and two other columns
which are complex conjugate to each other.
This can be explicitly proven as below.

The most general rotation in 3d Euclidean space which rotates the
space around an axis $\,\mathbf{n}=(n_1^{},\,n_2^{},\,n_3^{})^{T}$
by an angle $\,\phi\,$ is \cite{Rodejohann:2015pea},
%
\begin{equation}
R(\boldsymbol{n},\phi) \,=\,
\left(\begin{array}{ccc}
n_{1}^{2}\!+\! c\left(n_{2}^{2}\!+\!n_{3}^{2}\right)
& (1\!-\!c)n_1^{}n_2^{}\!+\! sn_3^{} & -sn_2^{}\!+\!(1\!-\!c)n_1^{}n_3^{}
\\[1mm]
(1\!-\!c)n_1^{}n_2^{}\!-\! sn_3^{} & c\!+\!n_{2}^{2}\!-\!cn_{2}^{2} &
sn_1^{}\!+\!(1\!-\!c)n_2^{}n_3^{}
\\[1mm]
sn_2^{}\!+\!(1\!-\!c)n_1^{}n_3^{} & -sn_1^{}\!+\!(1\!-\!c)n_2^{}n_3^{}
& c\!+\!n_{3}^{2}\!-\!cn_{3}^{2}
\end{array}\!\right) ,
\label{eq:1129}
\end{equation}
where $\,\mathbf{n}\cdot\mathbf{n}=1$\, and $\,(s,\,c)=(\sin\phi,\,\cos\phi)$.\,
We can directly verify that this matrix is diagonalized as
\begin{equation}
U_{R}^{\dag}R\,U_{R}^{} \,=\, \left(\begin{array}{ccc}
1 & 0 & 0    \\[0.5mm]
0 & c\!+\!\text{i}\hspace*{0.3mm} s & 0 \\[0.5mm]
0 & 0 & c\!-\!\text{i}\hspace*{0.3mm} s
\end{array}\!\right) ,
\label{eq:0708-6}
\end{equation}
where
\beqa
U_{R}^{} \,=\, \left(\,\,
\begin{matrix}
n_{1}^{} & -\frac{\sqrt{1-n_{1}^{2}}\,}{\sqrt{2}} & -\frac{\sqrt{1-n_{1}^{2}}\,}{\sqrt{2}}
\\[2mm]
n_{2}^{}
& \frac{n_1^{}n_2^{}-\text{i}\hspace*{0.3mm}n_3^{}}{\sqrt{2(1-n_{1}^{2})\,}}
& \frac{\,n_1^{}n_2^{}+\text{i}\hspace*{0.3mm}n_3^{}\,}{\sqrt{2(1-n_{1}^{2})\,}}
\\[3mm]
n_{3}^{} & \frac{n_{1}^{}n_{3}^{}+\text{i}\hspace*{0.3mm}n_{2}^{}}{\sqrt{2(1-n_{1}^{2})}} & \frac{\,n_1^{}n_3^{}-\text{i}\hspace*{0.3mm}n_2^{}\,}{\sqrt{2(1-n_{1}^{2})}\,}
\end{matrix}
\,\,\right) .
\label{eq:0708-7}
\eeqa
We see explicitly that the first column of $\,U_{R}^{}$ is real, and the second and third
columns are conjugate to each other, i.e., $\,U_R^\dag$\, has RCE.
Hence, if $\,R\in G_{\ell}^{}\,$,\, then $\,U_{\ell}^{\dag}=U_R^{\dag}$\, has RCE.


\end{appendix}

\vspace*{5mm}
\noindent
{\bf Acknowledgments}
\\[1.5mm]
We thank Gui-Jun Ding, Anjan S.\ Joshipura, C.\ S.\ Lam, Patrick Ludl, Ernest Ma,
Rabi Mohapatra, and Zhi-zhong Xing for useful discussions.
HJH is supported by the National NSF of China, and by the visiting grant
of IAS Princeton. WR is supported by the Max Planck Society in the
project MANITOP. XJX is supported by the China Scholarship Council (CSC).

\vspace*{3mm}
\noindent
{\bf Note added:}
While we were finalizing the present paper,
Ref.\,\cite{Joshipura:2015dsa} appeared on arXiv, which
has some partial overlap. It was also pointed out there that
$\mu$-$\tau$ reflection symmetry can be generated by
discrete residual subgroups of O(3).
In Sec.\,\ref{sub:reflection}, we explicitly showed that with
a proper basis transformation the $\mu$-$\tau$ reflection symmetry
is actually a real symmetry.
Our general theorems are independent and complementary to \cite{Joshipura:2015dsa},
and we presented a set of equivalent formulations for
the CMCPV as well as its parametrization-independent realization.



\end{document}